# Rapid and stain-free quantification of viral plaque via lens-free holography and deep learning

Tairan Liu[†,1,2,3], Yuzhu Li[†,1,2,3], Hatice Ceylan Koydemir[1,4,5], Yijie Zhang[1,2,3], Ethan Yang[1,6], Merve Eryilmaz[1,2], Hongda Wang[1,2,3], Jingxi Li[1,2,3], Bijie Bai[1,2,3], Guangdong Ma[1,7], and Aydogan Ozcan[*,1,2,3,8]

[1]Electrical and Computer Engineering Department, University of California, Los Angeles, CA, 90095, USA.
[2]Bioengineering Department, University of California, Los Angeles, 90095, USA.
[3]California NanoSystems Institute (CNSI), University of California, Los Angeles, CA, 90095, USA.
[4]Department of Biomedical Engineering, Texas A&M University, College Station, TX, 77843, USA.
[5]Center for Remote Health Technologies and Systems, Texas A&M University, College Station, TX, 77843, USA.
[6]Department of Mathematics, University of California, Los Angeles, CA, 90095, USA.
[7]School of Physics, Xi'an Jiaotong University, Xi'an, 710049, China.
[8]Department of Surgery, University of California, Los Angeles, CA, 90095, USA.

*Correspondence: Aydogan Ozcan Email: ozcan@ucla.edu

† Equal contributing authors

## Abstract

Plaque assay is the gold standard method for quantifying the concentration of replication-competent lytic virions. Expediting and automating viral plaque assays will significantly benefit clinical diagnosis, vaccine development, and the production of recombinant proteins or antiviral agents. Here, we present a rapid and stain-free quantitative viral plaque assay using lensfree holographic imaging and deep learning. This cost-effective, compact, and automated device significantly reduces the incubation time needed for traditional plaque assays while preserving their advantages over other virus quantification methods. This device captures ~0.32 Giga-pixel/hour phase information of the objects per test well, covering an area of ~30×30 mm$^2$, in a label-free manner, eliminating staining entirely. We demonstrated the success of this computational method using vesicular stomatitis virus (VSV), herpes simplex virus (HSV-1) and encephalomyocarditis virus (EMCV). Using a neural network, this stain-free device automatically detected the first cell lysing events due to the VSV viral replication as early as 5 hours after the incubation, and achieved >90% detection rate for the VSV plaque-forming units (PFUs) with 100% specificity in <20 hours, providing major time savings compared to the traditional plaque assays that take ≥48 hours. Similarly, this stain-free device reduced the needed incubation time by ~48 hours for HSV-1 and ~20 hours for EMCV, achieving >90% detection rate with 100% specificity. We also demonstrated that this data-driven plaque assay offers the capability of quantifying the infected area of the cell monolayer, performing automated counting and quantification of PFUs and virus-infected areas over a 10-fold larger dynamic range of virus concentration than standard viral plaque assays. This compact, low-cost, automated PFU quantification device can be broadly used in virology research, vaccine development, and clinical applications.

# Introduction

Viral infections pose significant global health challenges by affecting millions of people worldwide through infectious diseases, such as influenza, human immunodeficiency virus (HIV), human papillomavirus (HPV), and others[1]. The US Centers for Disease Control and Prevention (CDC) estimates that, since 2010, the influenza virus has resulted in 16-53 million illnesses, 0.2-1 million hospitalizations, and 16,700-66,000 deaths in the United States alone[2,3]. Furthermore, the ongoing COVID-19 pandemic has already caused >500 million infections and >6 million deaths worldwide, bringing a huge burden on public health and socioeconomic development[4]. To cope with these global health challenges, developing an accurate and low-cost virus quantification technique is crucial to clinical diagnosis[5], vaccine development[6], and the production of recombinant proteins[7] or antiviral agents[8,9].

Plaque assay was developed as the first method for quantifying virus concentrations in 1952 and was advanced by Renato Dulbecco, where the number of plaque-forming units (PFUs) was manually determined in a given sample containing replication-competent lytic virions[10,11]. These samples are serially diluted, and aliquots of each dilution are added to a dish of cultured cells[10]. As the virus infects adjacent cells and spreads, a plaque will gradually form, which can be visually inspected by an expert. Due to its unique capability of providing the *infectivity* of the viral samples in a *cost-effective* way, the plaque assay remains to be the gold standard method for quantifying virus concentrations despite the presence of other methods[12–19] such as the immunofluorescence focal forming assays (FFA)[14], polymerase chain reaction (PCR)[16], and enzyme-linked immunoassay (ELISA) based assays[19,20]. However, plaque assays usually need an incubation period of 2-14 days (depending on the type of virus and culture conditions)[21] to let the plaques expand to visible sizes, and are subject to human errors during the manual plaque counting process[22]. To improve the traditional plaque assays, numerous methods have been developed[23]. While these earlier systems have unique capabilities to image cell cultures in well plates, they require either fluorescence markers[22] or special culture plates with gold microelectrodes[24]. In addition, human counting errors still remain to be a problem for these methods. Hence, an accurate, quantitative, automated, rapid, and cost-effective plaque assay is urgently needed in virology research and related clinical applications.

Some of the recent developments in quantitative phase imaging (QPI), holography, and deep learning provide an opportunity to address this need. QPI is a preeminent imaging technique that enables the visualization and quantification of transparent biological specimens in a non-invasive and label-free manner[25,26]. Furthermore, the image quality of QPI systems can be enhanced using neural networks by improving e.g., phase retrieval[27], noise reduction[28], auto-focusing[29,30], and spatial resolution[31]. In addition, numerous deep learning-based microorganism detection and identification methods have been successfully demonstrated using QPI[32–42].

Here, we report a cost-effective and compact label-free live plaque assay that can automatically provide significantly faster quantitative PFU readout than traditional viral plaque assays without the need for staining. A compact lensfree holographic imaging prototype was built (Fig. 1 and Supplementary Video 1) to image the spatio-temporal features of the target PFUs during their incubation; the total cost of the parts of this entire imaging system is < $880, excluding a standard laptop computer. This lensfree holographic imaging system rapidly scans the entire area of a 6-well plate every hour (at a throughput of ~0.32 Giga-pixels per scan of a test well), and the reconstructed phase images of the sample are used for PFU detection based on the spatio-temporal changes observed within the wells. A neural network-based classifier was trained and used to convert the reconstructed phase images to PFU probability maps, which were then used to reveal the locations and sizes of the PFUs within the well plate. To prove the efficacy of our system, early detection of vesicular stomatitis virus (VSV), herpes simplex virus type 1 (HSV-1), and encephalomyocarditis virus (EMCV) were performed on Vero E6 cell plates. Our stain-free device could automatically detect the first cell-lysing event due to the VSV replication as early as 5 hours after the incubation and achieve >90% PFU detection rate in <20 hours, providing major time savings compared to the traditional plaque assays that take ≥48 hours. Furthermore, an average incubation time saving of ~48 hours and ~20 hours was demonstrated for HSV-1 and EMCV,

respectively, achieving a PFU detection rate >90% with 100% specificity. A quantitative relationship was also developed between the incubated virus concentration and the virus-infected area on the cell monolayer. Without any extra sample preparation steps, this deep learning-enabled label-free PFU imaging and quantification device can be used with various plaque assays in virology and might help to expedite vaccine and drug development research.

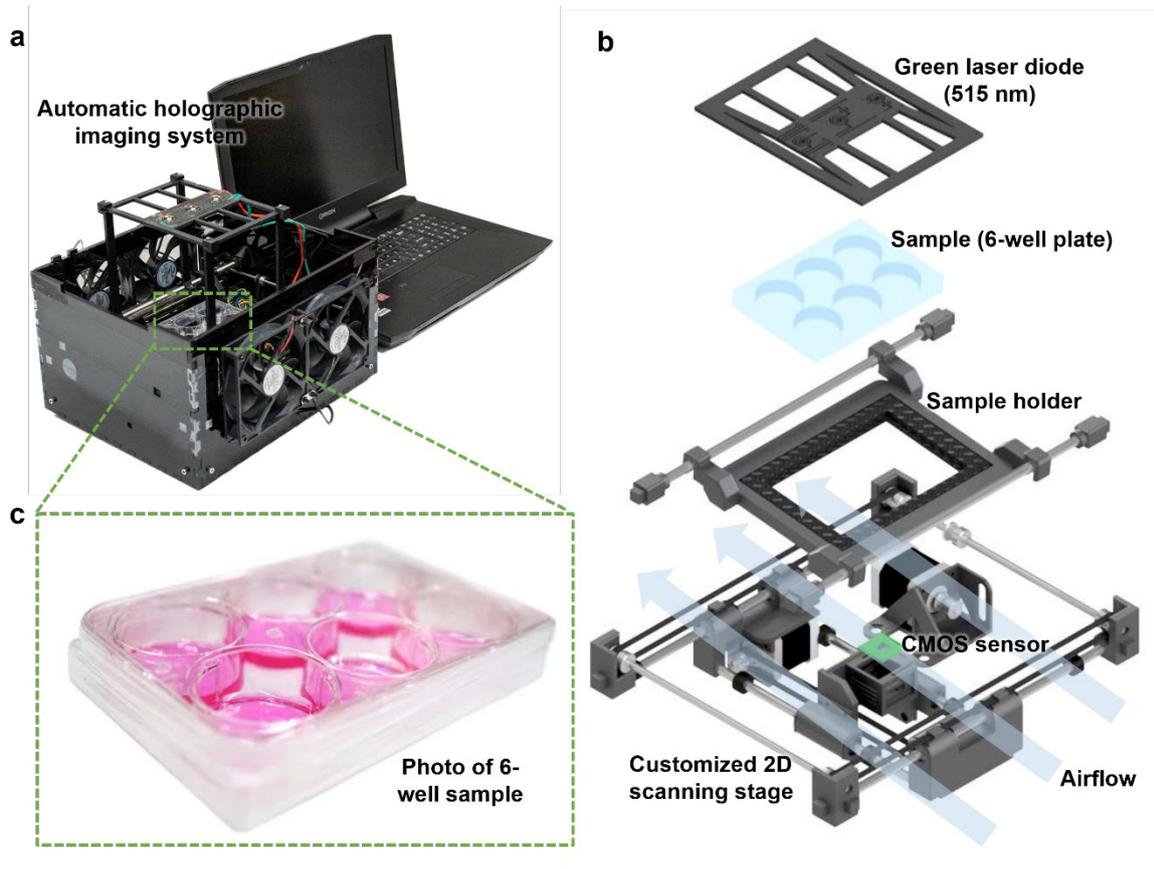

**Figure 1: Stain-free, rapid, and quantitative viral plaque assay using deep learning and lensless holography.** (a) Photograph of the stain-free PFU imaging system that captures the phase images of the plaque assay at a throughput of ~0.32 Giga-pixels per scan of each test well. The processing of each test well using the PFU classifier network takes ~7.5 min/well, automatically converting the holographic phase images of the well into a PFU probability map (see Fig. 2). (b) Detailed illustration of the system components. (c) A 6-well plate sample with ventilation holes on the cover and parafilm sealed from the side. Also see Supplementary Video 1.

## Results

To demonstrate the efficacy of the presented device, we prepared 14 plaque assays using the Vero E6 cells and VSV. The sample preparation steps followed standard plaque assays and are summarized in Fig. 2a (see the Methods section for details). For each 6 well-plate, ~$6.5 \times 10^5$ cells were seeded to each well, which was then incubated inside an incubator (Heracell™ VIOS 160i $CO_2$ Incubator, Thermo Scientific™) for 24 hours to achieve a cell monolayer with >95% coverage. During the virus infection, 5 wells were infected by 100 µL of the diluted VSV suspension (obtained by diluting a $6.6 \times 10^8$ PFU/mL VSV stock with a dilution factor of $2^{-1} \times 10^{-6}$), and 1 well was left for negative control. Then, 2.5 mL of the overlay solution containing the total medium with 4% agarose was added to each well (see the Methods section for details, subsection "Preparation of agarose overlay solution"). After the solidification of the overlay at room temperature, each sample was first placed into our imaging set-up for 20 hours of incubation, performing time-lapse imaging to capture the spatio-temporal information of the sample. Then, the same sample was left in the incubator for an additional 28 hours to let the PFUs grow to their optimal size for the traditional plaque assay (this is only used for comparison purposes). Finally, each sample was stained using crystal violet solution to serve as the ground truth to compare against our label-free method.

To train and test our network-based VSV PFU classifier, 54 wells (i.e., 45 positive wells and 9 negative wells) were used for training and 30 wells (i.e., 25 positive wells and 5 negative wells) were used for testing. During the training phase, a machine learning-based coarse PFU localization algorithm was developed to both accelerate the training dataset generation and delineate the potential false positives (see the Method sections for details). After this PFU localization algorithm screened each sample, the resulting PFU candidates were further examined manually for confirmation using a custom-developed Graphical User Interface (Supplementary Figure 1); this manual examination was only used during the training phase to correctly and efficiently prepare the training data. The negative training dataset was populated purely from the negative control well of each well plate. In total, 357 true positive PFU holographic videos and 1169 negative holographic videos were collected for training the PFU decision neural network. This dataset was further augmented to create a total of 2594 positive and 3028 negative holographic videos (see the Method sections), where each frame had 480×480 pixels, and the time interval between two consecutive holographic frames was 1 hour.

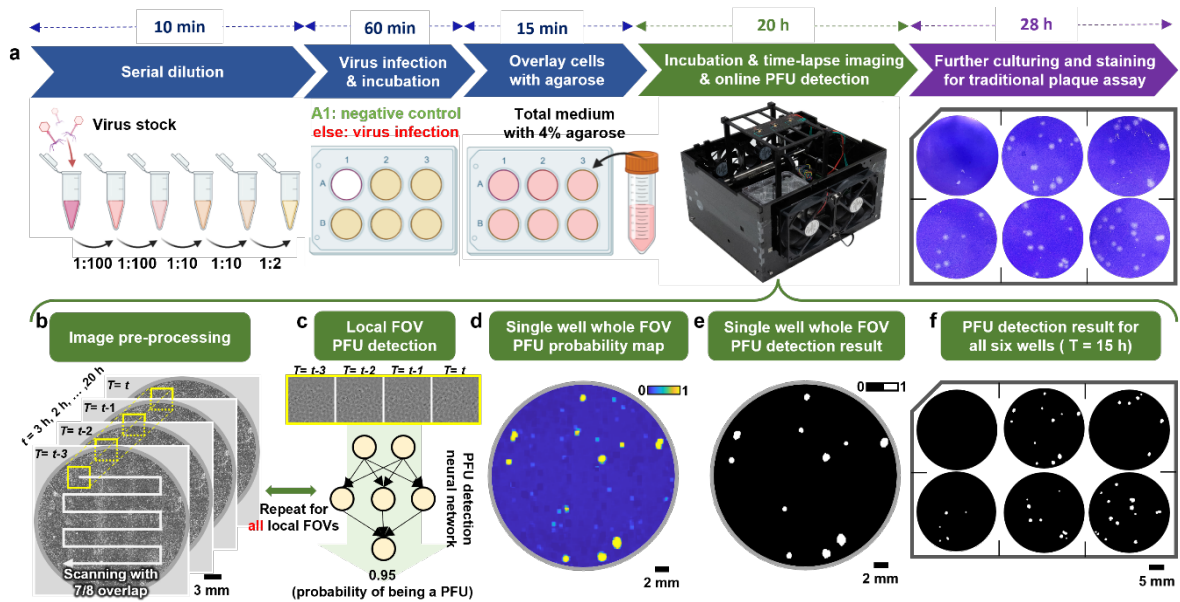

**Figure 2: Schematics of the workflow of the label-free viral plaque assay and its comparison to the standard PFU assay.** (a) Plaque assay sample preparation workflow. The traditional plaque assay at the last step in (a) is only performed for comparison purposes and is not needed for the operation of the presented PFU detection device. (b-f) Detailed image and data processing steps for the live viral plaque assay.

After the neural network-based VSV PFU classifier was trained, it was blindly tested on all 30 test wells in a scanning manner (shown in Fig. 2b) without the need for the PFU localization algorithm, which was only used for the training data generation. For each test well, we have ~18000×18000 effective pixels (representing a 30×30 $mm^2$ active area after discarding the edges); the digital processing of each test well using the PFU classifier network takes ~7.5 min, which automatically converts the holographic phase images of the well into a PFU probability map (Fig. 2d). Each pixel of the well on this map indicates the statistical probability of the local area (0.8×0.8 $mm^2$) centered at this pixel having a PFU. Using a probability threshold of 0.5, the final PFU detection and quantification result was obtained across the entire test well area (see e.g., Fig. 2 e-f). The impact of this probability threshold is analyzed and discussed in Supplementary Figure 2 and Supplementary Note 1, which illustrates the trade-off between the specificity and the sensitivity by selecting different threshold values.

Figure 3a shows examples of our device's performance in detecting VSV PFUs after 17 hours of incubation, representing a critical time that our detection rate exceeds 90% (Supplementary Figure 3 also shows our detection results after 15 hours and 20 hours of incubation, reported for comparison). Three representative PFUs are also selected and shown in Fig. 3b. When a PFU is in its early stage of growth, with its size much smaller than our 0.8×0.8 mm$^2$ virtual scanning window, it appears as a square (shown by the PFU① in Fig. 3b) in the final detection result, which effectively is the 2D spatial convolution of the small scale PFU with our scanning window. As another example, PFU③ shows a cluster forming event where the two neighboring PFUs can be easily differentiated using our method as opposed to the traditional plaque assay where they physically merged into one. Fig. 3c further shows the PFU quantification achieved by our device compared to the 48-hour traditional plaque assay results. We achieved a detection rate of >90% at 20 hours of incubation *without having any false positives at any time point* despite using no staining.

We also compared our results against a widely-adopted automatic PFU counting system that is commercially available. After the 48-hour incubation, followed by the standard staining protocol, we imaged the same five 6-well test plates (VSV, Fig. 3c) using this time the Agilent BioTek Cytation 5 device (Agilent Technologies, Santa Clara, CA). After the automated image acquisition with this system, the PFU detection was performed by Gen 5 software (Agilent Technologies, Santa Clara, CA) using the optimized settings of its automated PFU counting algorithm (see the Methods section). A detection rate of 94.3% was achieved with a 1.2% false discovery rate. In comparison, the presented stain-free holographic method achieved a PFU detection rate of 93.7% with 0% false discovery rate at 20 hours of incubation for the same samples (i.e., 28 hours earlier compared to the standard incubation time). In addition to missing some of the late-growing PFUs and introducing some false positives, this commercially available automated PFU counting system also showed over-segmentation on large PFUs and under-detection of PFUs for samples with high virus concentrations. A detailed report of the over-counted, false negative, and false positive PFUs, as well as a visualized PFU detection performance summary of this standard detection method compared to our device are demonstrated in Supplementary Figure 4.

In addition to saving incubation time and being stain-free, our presented framework also exhibits strong generalization capability. For example, after its training with 6-well plates, it can be directly used on 12-well plates without the need for any modifications or retraining steps (see e.g., the subsection "Well plate preparation" in the Methods section). Without any transfer learning steps, we achieved a PFU detection rate of 89% at 20 hours of incubation (VSV) when blindly tested on a 12-well plate (see Supplementary Figure 5). Furthermore, our computational PFU detection device can generalize to detect other types of viruses (e.g., HSV-1 and EMCV) through transfer learning while using the VSV PFU detection network as the base model. For HSV-1, two 6-well plates were prepared for transfer learning (see the Methods section), imaged for 72 hours with a 2-hour imaging interval/period, and further incubated for a total of 120 hours to obtain the stained ground truth PFU samples. The collected data were used to populate the training dataset for transfer learning. The resulting HSV-1 neural network was blindly tested on 12 additional HSV-1 test wells (containing in total 214 HSV-1 PFUs and 2 negative control wells); as shown in Supplementary Figure 6, without introducing false positives, our framework achieved 90.4% detection rate at 72 hours, reducing 48 hours of incubation time compared with the 120 hours required by the traditional HSV-1 plaque assay[43]. Similarly, for EMCV three 6-well plates were used for transfer learning (see the "Well plate preparation" subsection of the Methods), which were imaged for 60 hours with an imaging interval of 1 hour and stained at 72 hours of total incubation, following the standard protocols. When tested on 12 additional EMCV test wells (containing in total 249 EMCV PFUs and 2 negative control wells), a detection rate of 90.8% with 0% false positives was obtained at 52 hours of incubation, as shown in Supplementary Figure 7, achieving 20 hours of incubation time saving compared with the ground truth of 72 hours for the traditional EMCV plaque assay[44]. Notably, the EMCV plates contain much more late-growing PFUs compared to VSV or HSV-1, which is also in line with earlier observations[45]. The presented framework achieved a reliable EMCV plaque counting performance even for the PFU merging regions of a test well, as illustrated in Supplementary Figure 7c. Due to the spatio-temporal feature analysis-based early detection capability of our stain-free system, it could identify each individual PFU within these merging PFU regions at the early phases of the plaque growth, eliminating false negatives or misses that might have arisen in standard PFU counting methods due to the expansion of earlier PFUs, spatially covering (and obscuring) the late-growing plaques.

The presented device is cost-effective, compact, and automated, and can also handle a larger virus concentration range with a more reliable PFU readout. To demonstrate this, we prepared another 5 titer test plates, where for each plate, all 6 wells were infected by VSV, but with a 2 times dilution difference between each well, covering a large dynamic range in virus concentration from one test well to another. As shown in Fig. 4, our method is effective even for the higher virus concentration cases; see, for example, the dilution cases of $2^{-2} \times 10^{-4}$ and $2^{-3} \times 10^{-4}$. In the traditional 48-hour plaque assay, only the lowest virus concentration is suitable for the PFU quantification due to significant spatial overlapping, whereas for our label-free device, we can automatically and accurately count the individual PFUs at an early stage, even for the highest virus concentration (see Fig. 4c).

Furthermore, our method provides a more reliable readout; for example, in the circled region in Fig. 4 a-b, the absence of the cells was caused by some random cell viability problems that occurred during the plaque assay. In our device, these artifacts can be easily differentiated from the cell lysing events caused by the viral replication, since the spatio-temporal patterns for these two events are vastly different (assessed by the trained PFU probability network). This makes our deep learning-enabled device resilient to potential artifacts or cell viability issues randomly introduced during the sample preparation steps.

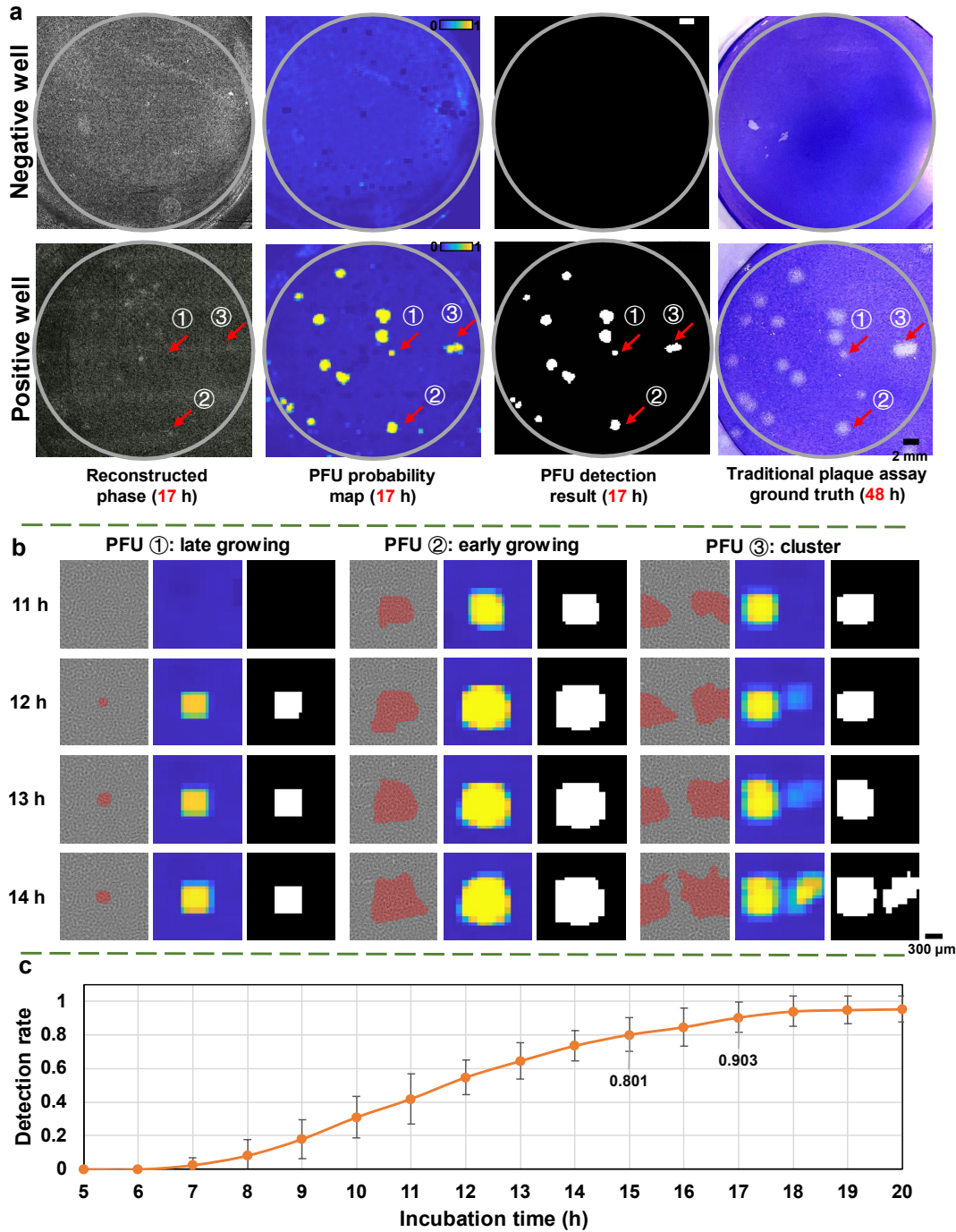

**Figure 3: Performance of the stain-free plaque assay for samples with low virus concentration.** (a) Whole well comparison of the stain-free viral plaque assay after 17-h incubation against the traditional plaque assay after 48-h incubation and staining. (b) The growth of three featured PFUs in the positive well from (a). The reconstructed phase channel is overlaid with the mask generated using the PFU localization algorithm to reveal their locations better. (c) Average PFU detection rate using the label-free viral plaque assay. The error bars show the standard deviation across the 5 testing plates.

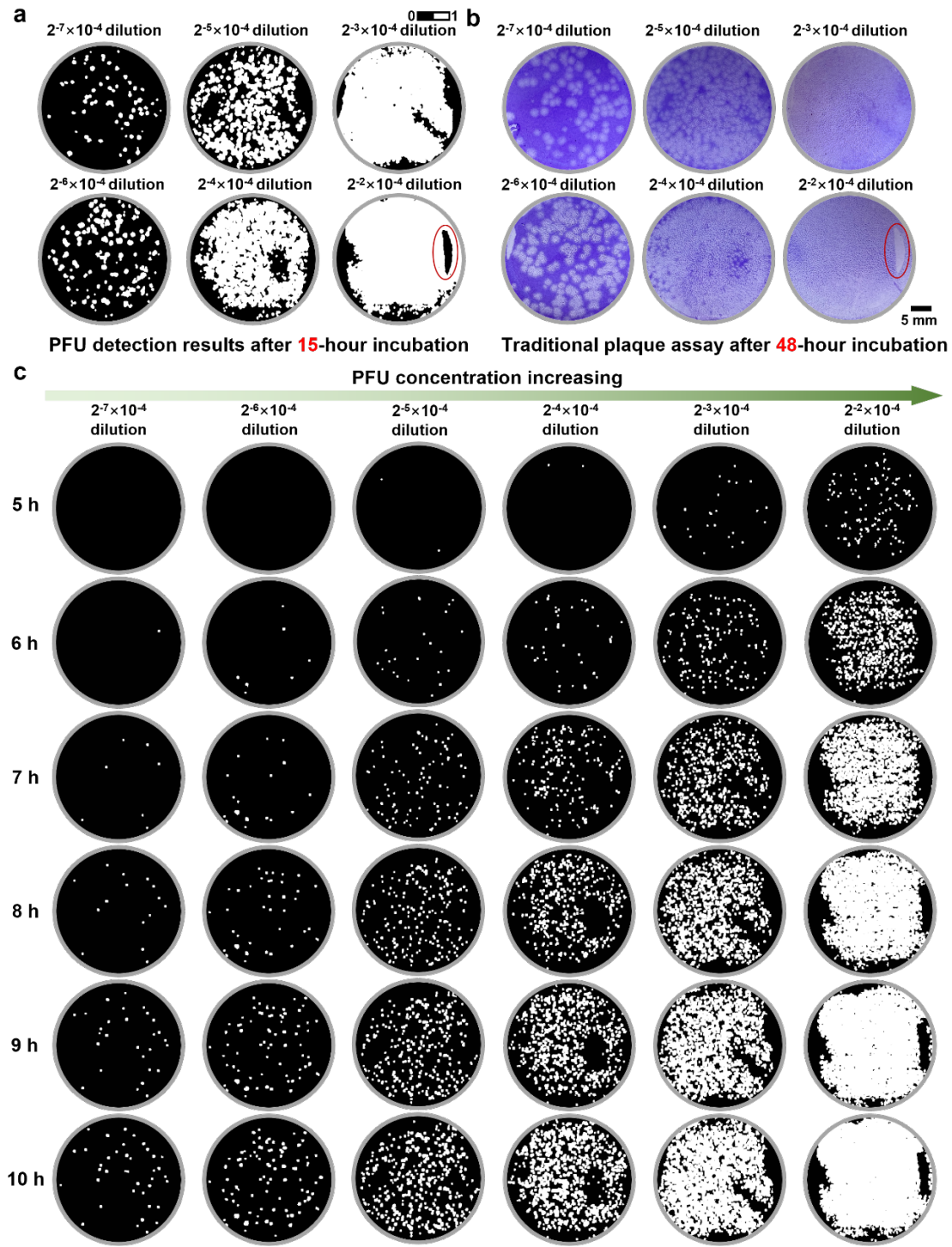

**Figure 4: Performance of the stain-free viral plaque assay as a function of the virus concentration.** (a-b) Whole plate comparison of the stain-free viral plaque assay after 15-h incubation against the traditional plaque assay after 48-h incubation and staining. (c) The growth of PFUs in their early stage for the same plate shown in (a) and (b).

Due to the high virus concentration used in these 5 titer test samples, PFUs quickly clustered and were no longer suitable for manual counting, as shown in Fig. 5a. However, the quantitative readout and the PFU probability map of our presented device allowed us to obtain the area of the virus-infected regions across all the time points during the incubation period, as shown in Fig. 5b. To better illustrate this, we plot in Fig. 5c the virus dilution factor vs. the ratio of the infected cell area per test well (in %) for all the samples at 6, 8, and 10 h of incubation time. Despite the existence of some serial dilution errors, late virus wakeups, and PFU clustering events, the infected area percentage that our device measured is monotonically decreasing with the increasing dilution factor for all the incubation times. This suggests that, by calibrating the system, the virus concentration (PFU/mL) can also be estimated from the percentage of the infected cell area per well.

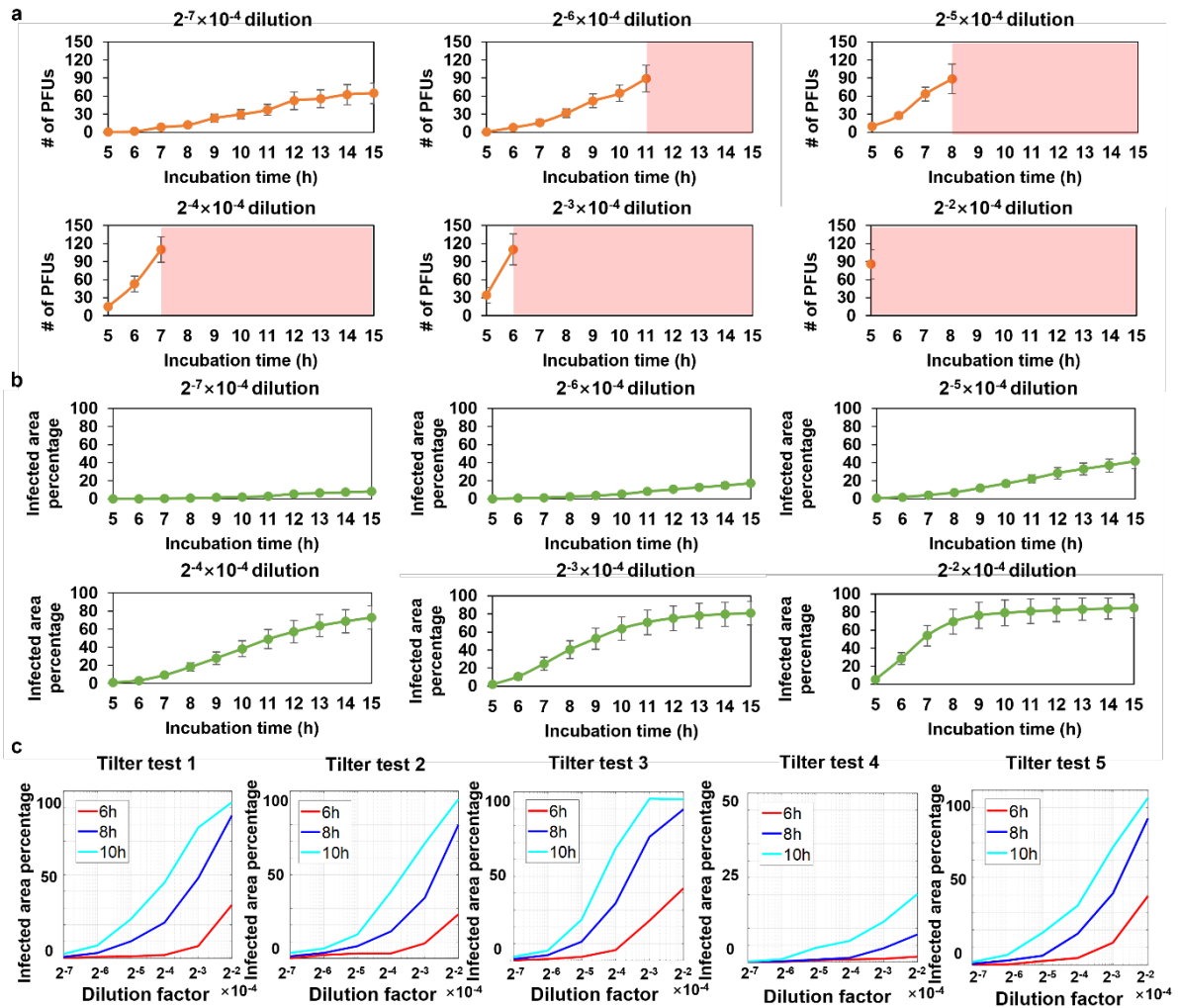

**Figure 5: Quantitative performance analyses of the label-free viral plaque assay for high virus concentration samples.** (a) PFU counting results for different high concentration virus samples at different time points. The light red region indicates the time when the PFUs were heavily clustered and no longer suitable for counting. (b) Area of the virus-infected regions for different high virus concentration samples at different time points. The error bars in (a-b) show the standard error across 5 titer testing plates. (c) Plots of virus dilution factor vs. the ratio of the infected cell area per test well (in %) for all 5 titer test samples at 6, 8, and 10 h of incubation time.

Furthermore, using the area percentage of the virus-infected region as a label-free quantification metric, the presented framework can provide earlier PFU readouts. To show this, we computed the infected area percentage for all the 25 positive/infected wells of the blind testing plates used to generate Fig. 3c. As shown in Fig. 6, when the infected area percentage is sufficiently large (>1%), a faster PFU concentration readout can be provided at 12-h or 15-h. Since the size of an average PFU on the well is physically larger at 15 hours of incubation compared to 12 hours, the slope of the red calibration curve in Fig. 6b is smaller than Fig. 6a, as expected. For samples with even higher virus concentrations, the infected cell area percentage could reach >1% in ⩽10 hours of incubation (shown in Fig. 5c), providing the PFU concentration readout even earlier.

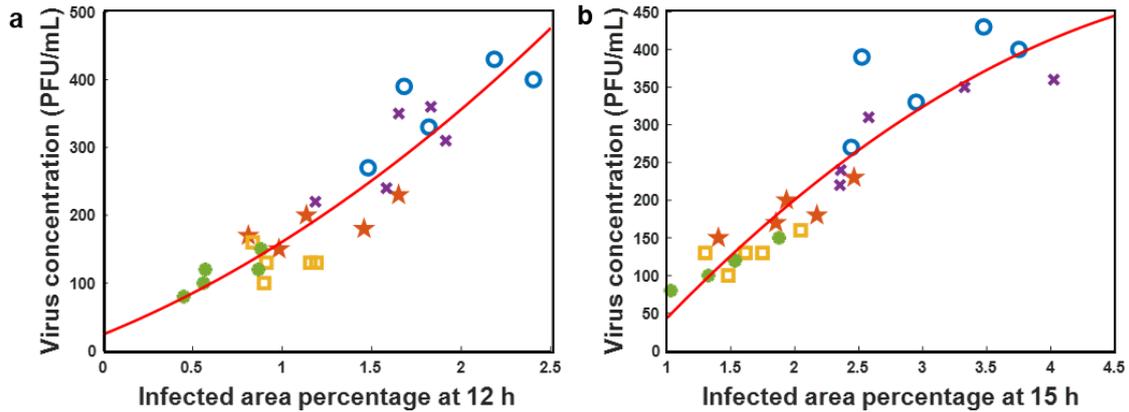

**Figure 6: Infected area percentage (%) measured by our stain-free device at different time points vs. the virus concentration per well (PFU/mL).** The virus concentrations in the *y*-axis were obtained from the 48-h traditional plaque assay for each test well. Different test wells of the same plate were marked with the same color/symbols. There are 25 infected test wells in each plot. The red calibration curves were obtained by quadratic polynomial fitting.

## Discussion

We demonstrated a cost-effective and automated early PFU detection system using a lensfree holographic imaging system and deep learning. This deep learning-based stain-free device captures time-lapse phase images of a test well at a throughput of ~0.32 Giga-pixels per scan, which is then processed by a PFU quantification neural network in ~7.5 min to yield the PFU distribution of each test well. The high detection rate of this label-free device with 100% specificity shown in Fig. 3c is a conservative estimate since the ground truth data were obtained after 48-h of incubation. In the early stages of the incubation period, many VSV PFUs did not even exist physically, which led to under-detection (e.g., a detection rate of 80.1% and 90.3% at 15 and 17 hours of incubation, respectively). This means that if we were to use the existing PFUs as the ground truth for our quantification at each time point, our detection rate would be even higher.

The success of this novel stain-free PFU detection system lies in the effective combination of digital holography and deep learning. The adoption of the lensfree holographic imaging system is essential for imaging unstained cells within a compact incubator, providing the spatio-temporal phase information of the samples using a compact, cost-effective and high-throughput imaging system. For a given time stamp of our imaging system, the PFU regions would in general express a wider phase distribution compared to the non-PFU regions; furthermore, a given PFU region would typically exhibit more dramatic phase changes across different time points (see Supplementary Figure 8 for some examples). These unique spatio-temporal signatures that are present in the phase channel of the holographic label-free time-lapse images are crucial for the deep neural network to statistically identify the target PFU regions from non-PFU regions at earlier time points, without introducing false positives or undercounting due to spatial overlaps. In addition, the large field-of-view (FOV) of the lensfree holographic on-chip imaging configuration with unit fringe magnification, along with its capability for digital focusing without any autofocusing hardware or objective lens helped us achieve a large phase information throughput of ~0.32 Giga-pixels in <30 sec per test well (covering a FOV of ~30×30 mm$^2$) using a compact and cost-effective device that can fit into any standard incubator without major modifications. This enabled us to rapidly scan an entire 6-well plate within 3 min, and as a result, our device can potentially scan the PFU samples even more frequently than every hour, which might enable further time savings in PFU detection using finer spatio-temporal changes that might be learned with a shorter imaging period. Such an approach would come with the trade-off of requiring significantly more training data and computation time.

Furthermore, due to the axial defocusing tolerance of our deep learning-based PFU detection method, the image reconstruction steps (spanning several hours of automated time-lapse imaging within an incubator) can be further simplified by propagating the acquired lensfree holograms to a fixed sample-to-sensor axial distance for the entire well *without* affecting the PFU detection results, while also ensuring a high throughput; for a demonstration of this, please see Supplementary Note 3 and Supplementary Figure 9 that quantify the defocusing distance tolerance of our system.

Moreover, our computational holographic PFU detection device requires negligible changes to the standard sample preparation steps employed in traditional plaque assays, while skipping the staining process entirely. The temperature, refractive index and optical field changes within the incubator caused by, e.g., evaporation, bubble formation etc., have negligible influence on the PFU detection performance of this system since such artifacts and statistical variations are learned during the training experiments, helping the trained neural networks successfully differentiate the spatio-temporal features of the true PFUs corresponding to viral replication from such fluctuations and physical perturbations within the incubator environment that naturally occur over several hours. Furthermore, our holographic time-lapse imaging system does not negatively influence or introduce a bias on the plaque formation process within the test wells, which is validated against control experiments as reported in Supplementary Figure 10.

The modular design employed by the presented PFU detection platform brings the potential for further system improvements. For example, parallel imaging can be achieved by installing several image sensors on the same system without significantly increasing the cost of the device, which will further improve the 30 $cm^2$/min effective imaging throughput of the device[46]. More accurate scanning stages can also help reduce the image registration steps needed during image pre-processing. Multi-wavelength phase recovery[47] can also be implemented to improve the overall image quality of the label-free plaques. The presented deep learning-enabled PFU detection framework can be potentially adapted to other imaging modalities that can provide the spatio-temporal differences in the PFU regions for various types of viruses; similarly, the trained PFU classifier network also has the adaptability to these system changes (see the Supplementary Note 2).

All in all, we presented a stain-free, rapid, and quantitative viral plaque assay using deep learning and holography. The presented compact and cost-effective device preserves all the advantages of the traditional plaque assays while significantly reducing the required sample incubation time in a label-free manner, saving time and eliminating staining. It is also resilient to potential artifacts during the sample preparation, and can automatically quantify a larger dynamic range of virus concentrations per well. We expect this technique to be widely used in virology research, vaccine development, and related clinical applications.

## Material and Methods

**Safety practices:** We handled all the cell cultures and viruses during our experiments at our biosafety level 2 (BSL2) laboratory according to the environmental, health, and safety rules and regulations of the University of California, Los Angeles. All operations were carried out under strict aseptic conditions.

**Studied organisms:** We used Vero C1008 [Vero 76, clone E6, Vero E6] (ATCC® CRL-1586™) (ATCC, USA), vesicular stomatitis virus (ATCC® VR-1238™), herpes simplex virus type 1 (ATCC® VR-260™) and encephalomyocarditis virus (ATCC® VR-129B™). Vero E6 cells are African green monkey kidney cells and are epithelial cells.

**Cell propagation:** We placed the frozen stock culture immediately in the liquid nitrogen vapor, until ready for use, just after the delivery of the frozen stock culture from ATCC. ATCC formulated Eagle's Minimum Essential Medium (EMEM) (product no. 30-2003, ATCC, USA) was used as a base medium for the cell line. For the complete growth medium, the base medium was mixed with fetal bovine serum (FBS) (product no. 30-2021, ATCC, USA) with a final concentration of 10 %. The FBS stock was aliquoted into 4 mL microcentrifuge tubes and stored at -20°C until use.

We used tissue culture flasks (75 $cm^2$ area, vented cap, TC treated, T-75) (product no. FB012937, Fisher

Scientific, USA) for cell culturing. The base medium in a T-75 flask and FBS were brought to 37°C in the incubator (product no. 51030400, ThermoFisher Scientific, Waltham, MA, USA) and fed with 5% $CO_2$ before handling it for cell culturing steps. The complete growth medium was prepared. The frozen cell culture was removed from liquid nitrogen and thawed under running water. After thawing the cells, the cell suspension was added to a T-75 flask containing 8 mL of complete growth medium (i.e., EMEM + 10% FBS). The flask was incubated at 37°C and 5% $CO_2$ in the incubator. The adherence of the cells to the flask surface was analyzed daily under a phase-contrast microscope. The medium in the flask was renewed 2-3 times a week. The cells were sub-cultivated in a ratio of 1:4 when 95% confluency of the cells as a monolayer was reached.

**Subculturing of cells:** After the removal of the medium from the cell culture flask, the cells were exposed to 2-3 mL of 0.25% Trypsin/0.53 mM EDTA (ATCC® 30-2101™, ATCC, USA) per flask for dissociation of cell monolayers. The flasks were kept in the incubator for 5-6 minutes for rapid dissociation of cells. 8 mL of complete medium per flask was added to each of them and 2-3 mL of the mixture containing suspended cells was transferred into a new T-75 flask. 8 mL of complete medium was added to the new flask and after gentle mixing, it was incubated at 37°C and 5% $CO_2$ for the growth of new cells.

**Virus propagation:** After the delivery of the virus stock samples from ATCC, they were stored in liquid nitrogen tanks until further use. Virus propagation requires to have Vero cells to be cultured and reach 90-95% confluency on the day of infection. Therefore, Vero cells were cultured for 1-2 days before the virus propagation using a seed cell suspension of Vero cells that were subcultured more than 3 times.

On the day of the virus infection, the growth medium in the Vero cell culture flask was removed and discarded. Then, it was rinsed using 5 mL Dulbecco's Phosphate Buffered Saline (D-PBS), 1X (ATCC® 30-2200™) (product no. 30-2200, ATCC, USA). After keeping the D-PBS containing flask for 3 min in the cabinet, the buffer solution was removed and discarded. For the virus propagation, the Vero cells in each flask were infected by 14 µL of VSV stock virus, 17 µL of HSV-1 stock virus, or 20 µL of EMCV stock virus with a multiplicity of infection (MOI) of 0.003, 0.07, and 0.05 for the VSV, HSV-1, and EMCV, respectively. Following this, 6 mL of EMEM (without FBS) was added to each flask. The flasks were incubated at 37°C for 1 hour and rocked at 15 min intervals to have a uniform spread of virus inoculum. After 1 hour, 10 mL of complete medium was added to each flask and the flasks were incubated at 37°C and 5% $CO_2$ for 48 h to 72 h.

After the incubation, the flasks were analyzed under a phase-contrast microscope. The cells should dissociate from the surface and round cells should be observed in the mixture if the virus propagation process is successful. The mixture was collected into a 50 mL tube (product no. 06-443-20, Fisher Scientific, USA) and the tubes were sealed using a parafilm layer. The suspension in the tube was centrifuged at ~2600 g for 10 min using a centrifuge with swing-out rotors (product no. 22500126, Fisher Scientific, USA). The supernatant containing the virus was collected from the tube and pooled in a new tube. After gentle mixing of the tube to have a uniform suspension, the suspension was aliquoted into 1 mL cryogenic vials with O-ring (product no. 5000-1012, Fisher Scientific, USA). The tubes were labeled and stored in liquid nitrogen tanks.

**Preparation of agarose solution:** 4% Agarose (product no. MP11AGR0050, Fisher Scientific, USA) in reagent grade water (product no. 23-249-581, Fisher Scientific, USA) was prepared and well mixed[48]. The suspension was then aliquoted into the glass bottles. The solution was sterilized at 121°C for 15 min in an autoclave and 50 mL aliquots were stored at 4°C until use.

**Preparation of agarose overlay solution:** One of the tubes containing the 50 mL of sterile agarose solution was heated up in a microwave oven for ~30 s. The solution was cooled down to 65°C in a water bath. 23.9 mL EMEM medium was mixed with 0.6 mL FBS and warmed to 50°C. 3.5 mL of agarose solution was added into the warmed medium mixture using a 10 mL- serological pipette and kept at 50°C until use.

**Well plate preparation:** First, the adhered cells in the flask were resuspended using trypsin. The solution was gently mixed to have uniform cell suspension and 10 μL of the suspension was taken for cell counting using a hemacytometer chamber. The cells were counted using a phase-contrast microscope. According to the cell count, the concentration of cells was adjusted to ~$6.5×10^5$ cells /mL by diluting the suspension using the complete medium. ~$6.5×10^5$ cells were added to each well of a new 6-well plate (product no. CLS5316, Corning, Glendale, AZ, USA). Then, 2 mL of complete medium was added to each well and the plate was stored at 37°C and 5% $CO_2$ for 24 h. Next, the cell coverage on each well was checked under the microscope. The cell coverage should reach ~95% to perform the PFU assay.

For a given 6-well plate, the cells of each well were infected with 100 μL of diluted virus suspension (the dilution factors for VSV, HSV-1, and EMCV are $2^{-1}×10^{-6}$, $2^{-2}×10^{-5}$, and $2^{-3}×10^{-3}$, respectively) and ~2.5-3 mL of the overlay solution was added to the cells. After the solidification of the overlay at room temperature, the plate was incubated in an incubator (Heracell™ VIOS 160i $CO_2$ Incubator, Thermo Scientific™) for 48 hours, 120 hours and 72 hours corresponding to VSV, HSV-1, and EMCV, respectively. A photo comparison of the HSV-1 samples at 72 hours, 96 hours and 120 hours of incubation is shown in Supplementary Figure 11, which confirms the need for 120 hours of incubation for HSV-1 PFUs. Similarly, a photo comparison of the EMCV samples at 48 hours and 72 hours of incubation is shown in Supplementary Figure 12, confirming the need for 72 hours of incubation for EMCV. These observations are also in line with previous studies[43,44].

The preparation of the 12-well plates used for VSV PFU testing followed the same workflow of the 6-well plate VSV preparation. The only difference in preparing 12-well VSV plates is that the seeded cells in each well, the virus suspension volume per well, and the agarose overlay solution used for each well were reduced to half compared with the 6-well plates. We summarized the different experimental settings that were used for VSV, HSV-1, and EMCV in the process of virus propagation and well plate preparation in Supplementary Table 2.

**Preparation of crystal violet solution:** 0.1 g of crystal violet powder (product no. C581-25, Fisher Scientific) was mixed with 40 mL reagent grade water in a 50 mL centrifuge tube. The mixture was gently mixed to dissolve the powder. 10 mL methanol (product no. A452-4, Fisher Scientific) was added to the mixture and stored at room temperature.

**Fixation and staining of cells:** These steps were only performed for comparison against our device's PFU readings. After 48 h of VSV incubation, 120 h of HSV-1 incubation, or 72 h of EMCV incubation, the plate was removed from the incubator and the cells were fixed using 0.5 mL methanol/acetic acid solution for 30 min. After 30 min, the wells were washed with water gently to remove the agarose layer. The excess water was removed, and 1 mL of crystal violet (CV) solution was added to each well. The plate with CV was placed into the shaker incubator and mixed at 100 rpm for 30 min. After 30 min of incubation, tap water was used to remove excess stain from the plate and the waste was collected into a large beaker. The plate was left to dry in a fume hood and stored at room temperature by covering with an aluminum foil.

**Lens-free imaging set-up:** An automatic lens-free PFU imaging set-up was built to capture the in-line holograms of the samples. This set-up includes: 1) a holographic imaging system, 2) a 2D mechanical scanning platform, 3) a cooling system, 4) a controlling circuit, and 5) an automatic controlling program. Three green laser diodes (at 515 nm, 2 nm bandwidth, 0.17 mm emission diameter, Osram PLT5510) were used for coherent illumination, where each laser diode illuminates two wells on the same column of the 6-well sample plate (see Supplementary Video 1). The laser diodes were controlled by a driver (TLC5916, Texas Instruments, Texas, US) and mounted ~16 cm away from the sample. A CMOS image sensor (acA3800-14 μm, Basler AG, Ahrensburg, Germany, 1.67 μm pixel size, 6.4 mm × 4.6 mm FOV) was placed ~5 mm beneath the sample forming a lensfree holographic imaging system. The phase changes in the PFU regions were encoded in the acquired holograms.

There are several factors that affect the spatial resolution of the lensfree holographic imaging system, including 1) the spatial coherence of the illumination; 2) the temporal coherence of the illumination; 3) axial distance between the source aperture and the sample plane (referred to as $z_1$) and the sample-to-sensor plane distance ($z_2$); and 4) pixel size of the image sensor. As for the illumination source per well, we used a single-mode laser diode with a core size of 9 µm, with $z_1 \approx 16$ cm between the source plane and the sample plane, which provided sufficient spatial coherence covering the entire sample plane per well. As for the temporal coherence length of our illumination source, we have:

$$\Delta L_c \approx \sqrt{\frac{2\ln 2}{\pi n}} \cdot \frac{\lambda^2}{\Delta \lambda} = 88.09 \text{ µm} \qquad (1)$$

where, $\lambda = 515$ nm and $\Delta\lambda = 2$ nm, which is the bandwidth of the laser diode. We can accordingly calculate the effective numerical aperture due to the temporal coherence limit of the illumination light as ($NA_{temporal}$):

$$NA_{temporal} = n\sin\theta_{temporal} = n\sqrt{1-\cos^2\theta_{temporal}} = n\sqrt{1-\left(\frac{z_2}{z_2+\Delta L_c}\right)^2} \approx 0.1853 \qquad (2)$$

where $z_2 \approx 5$ mm. This temporal coherence-based NA is lower than the effective numerical aperture that is dictated by the sample-to-sensor distance and the extent of the detector plane, and therefore, the temporal coherence-dictated holographic resolution limit of our system can be approximated as:

$$d_{coherence} \propto \frac{\lambda}{NA_{temporal}} = 2.7793 \text{ µm} \qquad (3)$$

Since our holographic on-chip imaging system has $z_1 \gg z_2$, it operates under a unit fringe magnification[49] and the native pixel size (1.67 µm) at the sensor plane also casts its own resolution limit due to the pixelation of the acquired holograms, unless pixel super-resolution[50,51] (PSR) approaches are utilized to digitally reduce the effective pixel size of each holographic frame. In this work, PSR was not utilized as our device acts as a PFU detector by sensing the spatio-temporal changes induced by viral replication events, and therefore a high spatial resolution (e.g., <1-2 µm) reconstruction of holograms was not necessary. In fact, these design choices also helped us significantly simplify and speed up the image processing pipeline and eliminate unnecessary data acquisition. Furthermore, the numerical spatio-temporal variations that might be introduced due to pixel super-resolution algorithms as a function of the incubation time might have introduced technical challenges for the learning of the PFU classifier neural networks, which is another design consideration that we had in addition to the simplification of the holographic data acquisition, processing and storage.

The FOV of the CMOS image sensor is ~0.3 cm², hence mechanical scanning is needed for imaging the whole area of a 6-well plate. A scanning platform was built using a pair of linear translation rails, a pair of linear bearing rods, and linear bearings. 3D printed parts were also used to aid with housing and joints. Two stepper motors (product no. 1124090, Kysan Electronics, San Jose, CA, USA), driven by two driver chips (DRV8834, Pololu Las Vegas, NV, US), were exploited to enable the CMOS sensor to perform 2D horizontal movement. This low-cost platform carries the CMOS sensor moving in a raster pattern and images a total of 420 holograms (21 horizontal, 20 vertical, with 15% overlap) in ~3 min to complete the whole sample scanning (see Supplementary Video 1).

The selected CMOS sensor could heat up to >70°C during its operation, which could disturb the growth of the sample and vaporize the agarose layer, especially for regions that are near the sensor parking location between successive holographic scans. Hence, a cooling system was built using fans (QYN1225BMG-A2, Qirssyn, China). We also sealed the sides of the sample using parafilm (product no. 13-374-16, Fisher Scientific, Hampton, NH, USA) and opened 4 holes on the top cover to form a gentle ventilation system, which is an inexpensive and easy-to-implement solution to avoid sample drying.

A microcontroller (Arduino Micro, Arduino LLC) was used to control the two stepper motor driver chips, the illumination driver chip, and a field-effect transistor-based digital switch (used to turn the CMOS sensor on/off). All these chips along with the digital switch, wires, and capacitors, were integrated on one printed circuit board (PCB), powered by a 6V-1A power adaptor connected to the wall plug.

An automatic controlling program with a graphical user interface (see Supplementary Figure 13) was developed using the C++ programming language. It can be used to adjust the image capture parameters (e.g., exposure time etc.) of the CMOS image sensor and communicate with the microcontroller to further switch the laser diodes or CMOS sensor on/off and control the movement of the mechanical scanning system.

All the components along with their unit prices are also summarized in Supplementary Table 1. The cost of the parts of this entire imaging system is < $880, excluding the laptop computer. At higher volumes of manufacturing, this cost can be significantly reduced.

**Image pre-processing:** After the image acquisition for each time interval, the raw holograms were firstly reconstructed using the angular spectrum approach based on back-propagation[49,52–55]. The accurate sample-to-sensor distance was estimated at the central region of each well using an auto-focusing algorithm based on the Tamura-of-Gradient (ToG) metric[56]. The same sample-to-sensor distance was used for the entire well since the neural network-based method can tolerate de-focusing. Then, the phase channel of the reconstructed holograms was stitched into the whole FOV image using a correlation-based method and linear blending[32].

Starting from the second time interval, a 2-step image registration was performed to compensate for the low accuracy of the mechanical scanning stage. A coarse whole FOV correlation-based image registration was firstly performed, then a local fine elastic image registration was followed[57]. The impact of this 2-step image registration is shown in Supplementary Video 2.

**Coarse PFU localization algorithm**

First, each current frame was stacked with the previous 3 frames (shown in Supplementary Figure 14a) and a background image (shown in Supplementary Figure 14b) was estimated through singular value decomposition[58]. By subtracting this background image, signals from the static regions were suppressed (shown in Supplementary Figure 14c). Then, by applying bilateral filtering, the PFU regions with high spatial frequency features were further enhanced (shown in Supplementary Figure 14d).

93 image patches (256 × 256 pixels) in PFU regions and 93 image patches from non-PFU regions were cropped manually from 3 experiments. Each pixel of these image patches was labeled as 1 for the PFU region and 0 for the non-PFU region. A Naïve Bayes pixel-wise classifier was trained using this dataset, where the Tamura-of-Gradient (ToG) metric[56] was computed at 2×, 4×, 8×, 16×, and 32× down-sampling scales to serve as the manually selected features. The effect of this classifier is shown in Supplementary Figure 14e. Finally, by applying several morphological operations (such as image close, image fill, etc.), the PFU regions are coarsely located (shown in Supplementary Figure 14f).

Though this coarse PFU localization algorithm was still subject to detecting false positives (shown in Supplementary Figure 14g), it could significantly simplify the effort needed for populating the network training dataset. In addition, applying this algorithm to a negative well would help delineate the potential false positives during network training (shown in Supplementary Figure h). Important to note that this PFU localization algorithm was only used for the training data generation, and was not employed in the blind testing phase as its function was to streamline the training data generation process to be more efficient.

**Network training dataset**

The network training datasets used in our work were generated by combining the coarse PFU localization algorithm with human labeling. To obtain the training datasets for VSV, 54 training wells from 9 6-well plates containing 9 negative control wells and 45 positive (virus-infected) wells were imaged and processed. For the positive training dataset, after the image pre-processing, the coarse PFU localization algorithm was applied to the images obtained at 12 hours of incubation. From the 45 positive wells, this process automatically generated 6930 VSV PFU candidates. Then, each of these candidates was examined by 4 experts using the customized Graphical User Interface shown in Supplementary Figure 2. Only those PFU candidates confirmed by all 4 experts were kept in the positive training dataset; potentially missed PFUs are not a concern here since this is just the training dataset. Ultimately, 357 positive videos of the confirmed PFUs were kept and were further populated to 2594 videos by performing augmentation over time. For the negative training dataset, all the negative videos were populated from the 9 negative control wells. To enhance the specificity of the network, the coarse PFU localization algorithm was also applied to the holographic images obtained at 12 hours of incubation. Any detected PFU regions were false positives in this case since these were from the negative control wells. However, such regions might contain unique spatial-temporal features that would potentially confuse the PFU network and thus were kept in the negative training dataset to provide valuable training examples for our deep neural network. In total, 1169 such videos were found by this process, and the negative training dataset is further augmented to 3028 videos by random selection from the negative control wells. Following the same dataset generation method, the training datasets of HSV-1 and EMCV that were used for transfer learning were prepared accordingly. The above-mentioned coarse PFU localization algorithm was first applied to 72-hour holographic phase images for HSV-1 and 60-hour holographic phase images for EMCV. For the HSV-1 training dataset, 1058 positive videos of 122 confirmed HSV-1 PFUs from 10 wells, and 1453 negative videos from 2 negative control wells were generated. Similarly, 776 positive videos of 152 EMCV PFUs from 15 wells and 1875 negative videos from 3 negative control wells formed the training dataset for EMCV. Based on the plaque-forming speed for each type of virus, the time intervals between 2 consecutive holographic frames for the VSV videos, HSV-1 videos and EMCV videos were set to 1 hour, 2 hours and 1 hour, respectively.

**Network architecture and training schedule**

Our PFU classifier network was built based on the DenseNet[59] structure, with 2D convolution layers replaced by the pseudo-3D building blocks[60]. The detailed architecture is shown in Supplementary Figure 15 and described in Supplementary Note 3. We used ReLU as the activation function. Batch normalization and dropout with a rate of 0.5 were used in the training. The loss function we used was the weighted cross-entropy loss:

$$l(p,g) = \sum_{k=1}^{K} \sum_{l=1}^{2} -w_l \cdot g_{k,l} \cdot \log\left( \frac{\exp(p_{k,l})}{\exp(p_{k,1}) + \exp(p_{k,2})} \right) \quad (4)$$

where $p$ is the network output, which is the probability of each class (i.e., PFU or non-PFU) before the SoftMax layer, $g$ is the ground-truth label (which is equal to 0 or 1 for binary classification), $K$ is the total number of training samples in one batch, $w$ is the weight assigned to each class, defined as $w = 1-d$, where $d$ is the percentage of the samples in one class ($d = 46.1\%$ for positive class, $d = 53.9\%$ for negative class).

The input 4-frame videos were formatted as a tensor with the dimension of $1 \times 4 \times 480 \times 480$ (channel × time frame × height × width). Data augmentation, such as flipping, and rotation were applied when loading the training dataset. The network model was optimized using the Adam optimizer with a momentum coefficient of (0.9, 0.999). The learning rate started as $1\times10^{-4}$ and a scheduler was used to decrease the learning rate with a coefficient of 0.7 at every 30 epochs. Our model was trained for 264 epochs using NVIDIA GeForce RTX3090 GPU with a batch size of 30. The loss curve, training sensitivity, and specificity curves of our training process are provided in Supplementary Figure 16. In these curves, 10% of the training dataset was randomly selected as the validation dataset. Note that the training and validation datasets (containing holographic videos of the wells) were formed from various wells at different time points of each PFU assay as detailed earlier; therefore, these training and validation sensitivity and specificity curves do not reflect the evaluation of an individual test well that is periodically monitored from the beginning of the incubation. Our blind testing results reported in the Results section, however, were acquired by using the trained VSV PFU detection neural network on individual test wells that were continuously monitored from the beginning of the incubation with a sampling period of 1 hour, achieving >90% detection rate for VSV PFUs with 100% specificity in <20 hours.

Similarly, we built the PFU detection neural networks for HSV-1 and EMCV through transfer learning, where the same neural network architecture was used, but initialized with the parameters obtained by the previously trained VSV model. Other training settings for HSV-1 and EMCV models, such as the loss function, initial learning rate, and optimizer, were all kept as same as the VSV model, but the learning rate was decreased with a coefficient of 0.8 every 10 epochs. Finally, the HSV-1 and EMCV models were obtained after 135 epochs and 88 epochs of training, respectively, based on the validation loss.

**Image post-processing**

After getting the PFU probability map and applying the 0.5 threshold, two image post-processing steps were followed to obtain the final PFU detection result: 1) maximum probability projection along time, and 2) PFU size thresholding. The maximum projection was used to compensate for the lower PFU probability values generated from the PFU center when it enters the late stage of its growth. The effect of this maximum projection is illustrated in Supplementary Figure 17. The size threshold on the PFU probability map was set to $0.5 \times 0.5$ mm$^2$.

**Automated PFU counting algorithm**

After getting the binary PFU detection mask for each test well, an automated PFU counting algorithm that is compatible with both sparse and dense viral samples was developed. First, the connected components in the detection mask at the m-th hour (denoted as $D_m$) were found. Then, the PFU counts for each connected component in $D_m$ were calculated by taking the maximum number of connected components that emerged in this region over time:

$$n_{cc} = \max_{t=[1,m]} \left( H\left( D_t * C \right) \right) \tag{5}$$

where $n_{cc}$ means the PFU count for the examined connected component in $D_m$, $D_t$ means the PFU detection mask at $t$-th hour, $C$ represents a binary map (with the same dimensions as $D_t$) which only maintains the current examined connected component in $D_m$ as 1, * means the element-wise multiplication, and $H(\cdot)$ refers to the operation of taking the number of the connected components. Finally, the sum of the $n_{cc}$ for all the connected components in $D_m$ was taken as the final PFU count for each well.

**Automated PFU counting settings for Biotek Citation 5**

For comparison against our device, some of the VSV 6-well plates were analyzed using the Biotek Citation 5 (Agilent Technologies, Santa Clara, CA) under the bright-field mode with an objective lens of 4× 0.13 NA. These captured images were processed and analyzed using its self-contained Gen5 Image Prime software (Agilent Technologies, Santa Clara, CA). The captured local images were first stitched into a whole FOV image of each test well, which was then processed by the "digital phase contrast" function using a 50 μm structuring element size. Next, the "cellular analysis" tool was used to perform the automated PFU counting. In its basic settings, an intensity threshold of 2500 and an object size threshold of 1500-5000 μm were used. And in its advanced detection settings, the rolling ball diameter of the background flattening, image smoothing strength, and the evaluated background level were set to 1000 μm, 20 cycles of 3×3 average filter, and 30% of the lowest pixels, respectively. All the parameters used for pre-processing and automated PFU counting were optimized in consultation with the technical support team from Agilent Technologies.

**PFU detection rate and the false discovery rate**

To evaluate the PFU detection performance of our device, the detection rate and the false discovery rate were defined as follows:

$$\text{Detection rate} = \frac{\text{TP}}{\text{GT}} \qquad (6)$$

where TP (true positives) represents the number of the detected PFUs by our device at a given time point within the incubator; GT (ground truth) is the total PFU number counted by an expert for the same sample after 48 hours of VSV incubation (120 hours for HSV-1 and 72 hours for EMCV) followed by the standard staining as part of the traditional plaque assay protocol. We also used:

$$\text{False discovery rate} = \frac{\text{FP}}{\text{TP + FP}} \qquad (7)$$

where FP stands for false positives.

## References

1. Singh, L., Kruger, H. G., Maguire, G. E. M., Govender, T. & Parboosing, R. The role of nanotechnology in the treatment of viral infections. *Ther Adv Infect Dis* **4**, 105–131 (2017).

2. CDC. Burden of Influenza. *Centers for Disease Control and Prevention* https://www.cdc.gov/flu/about/burden/index.html (2022).

3. CDC. Preliminary In-Season 2021-2022 Flu Burden Estimates. *Centers for Disease Control and Prevention* https://www.cdc.gov/flu/about/burden/preliminary-in-season-estimates.htm (2022).

4. COVID-19 Map. *Johns Hopkins Coronavirus Resource Center* https://coronavirus.jhu.edu/map.html.

5. Ryu, W.-S. Diagnosis and Methods. *Molecular Virology of Human Pathogenic Viruses* 47–62 (2017) doi:10.1016/B978-0-12-800838-6.00004-7.

6. Wen, Z. *et al.* Development and application of a higher throughput RSV plaque assay by immunofluorescent imaging. *Journal of Virological Methods* **263**, 88–95 (2019).

7. Basak, S., Kang, H.-J., Chu, K.-B., Oh, J. & Quan, F.-S. Simple and rapid plaque assay for recombinant baculoviruses expressing influenza hemagglutinin. *Science Progress* **104**, 00368504211004261 (2021).

8. Abou-Karam, M. & Shier, W. T. A Simplified Plaque Reduction Assay for Antiviral Agents from

Plants. Demonstration of Frequent Occurrence of Antiviral Activity in Higher Plants. *J. Nat. Prod.* **53**, 340–344 (1990).

9. Stepp, P. C. & Ph.D. New Method for Rapid Virus Quantification. *GEN - Genetic Engineering and Biotechnology News* https://www.genengnews.com/magazine/144/new-method-for-rapid-virus-quantification/ (2010).

10. Ryu, W.-S. *Molecular virology of human pathogenic viruses*. (Academic Press, an imprint of Elsevier, 2017).

11. Mendoza, E. J., Manguiat, K., Wood, H. & Drebot, M. Two Detailed Plaque Assay Protocols for the Quantification of Infectious SARS-CoV-2. *Current Protocols in Microbiology* **57**, cpmc105 (2020).

12. Rashid, K. A., Hevi, S., Chen, Y., Cahérec, F. L. & Chuck, S. L. A Proteomic Approach Identifies Proteins in Hepatocytes That Bind Nascent Apolipoprotein B *. *Journal of Biological Chemistry* **277**, 22010–22017 (2002).

13. Blaho, J. A., Morton, E. R. & Yedowitz, J. C. Herpes Simplex Virus: Propagation, Quantification, and Storage. *Current Protocols in Microbiology* **00**, 14E.1.1-14E.1.23 (2006).

14. Cruz, D. J. M. & Shin, H.-J. Application of a focus formation assay for detection and titration of porcine epidemic diarrhea virus. *J Virol Methods* **145**, 56–61 (2007).

15. Loret, S., El Bilali, N. & Lippé, R. Analysis of herpes simplex virus type I nuclear particles by flow cytometry. *Cytometry Part A* **81A**, 950–959 (2012).

16. Gallaher, S. D. & Berk, A. J. A rapid Q-PCR titration protocol for adenovirus and helper-dependent adenovirus vectors that produces biologically relevant results. *J Virol Methods* **192**, 28–38 (2013).

17. Killian, M. L. Hemagglutination Assay for Influenza Virus. in *Animal Influenza Virus* (ed.

Spackman, E.) 3–9 (Springer, 2014). doi:10.1007/978-1-4939-0758-8_1.

18. Roingeard, P., Raynal, P., Eymieux, S. & Blanchard, E. Virus detection by transmission electron microscopy: Still useful for diagnosis and a plus for biosafety. *Rev Med Virol* **29**, e2019 (2019).

19. Alhajj, M. & Farhana, A. Enzyme Linked Immunosorbent Assay. in *StatPearls* (StatPearls Publishing, 2022).

20. Pankaj, K. Virus Identification and Quantification. *Materials and Methods* (2021).

21. Baer, A. & Kehn-Hall, K. Viral Concentration Determination Through Plaque Assays: Using Traditional and Novel Overlay Systems. *J Vis Exp* 52065 (2014) doi:10.3791/52065.

22. Masci, A. L. *et al.* Integration of Fluorescence Detection and Image-Based Automated Counting Increases Speed, Sensitivity, and Robustness of Plaque Assays. *Mol Ther Methods Clin Dev* **14**, 270–274 (2019).

23. Ke, N., Wang, X., Xu, X. & Abassi, Y. A. The xCELLigence System for Real-Time and Label-Free Monitoring of Cell Viability. in *Mammalian Cell Viability: Methods and Protocols* (ed. Stoddart, M. J.) 33–43 (Humana Press, 2011). doi:10.1007/978-1-61779-108-6_6.

24. Burmakina, G., Bliznetsov, K. & Malogolovkin, A. Real-time analysis of the cytopathic effect of African swine fever virus. *Journal of Virological Methods* **257**, 58–61 (2018).

25. Park, Y., Depeursinge, C. & Popescu, G. Quantitative phase imaging in biomedicine. *Nature Photon* **12**, 578–589 (2018).

26. Cacace, T., Bianco, V. & Ferraro, P. Quantitative phase imaging trends in biomedical applications. *Optics and Lasers in Engineering* **135**, 106188 (2020).

27. Rivenson, Y., Zhang, Y., Günaydın, H., Teng, D. & Ozcan, A. Phase recovery and holographic image reconstruction using deep learning in neural networks. *Light Sci Appl* **7**, 17141–17141 (2018).

28. Choi, G. *et al.* Cycle-consistent deep learning approach to coherent noise reduction in optical diffraction tomography. *Opt. Express, OE* **27**, 4927–4943 (2019).

29. Luo, Y., Huang, L., Rivenson, Y. & Ozcan, A. Single-Shot Autofocusing of Microscopy Images Using Deep Learning. *ACS Photonics* **8**, 625–638 (2021).

30. Ding, H. *et al.* Auto-focusing and quantitative phase imaging using deep learning for the incoherent illumination microscopy system. *Opt. Express, OE* **29**, 26385–26403 (2021).

31. Liu, T. *et al.* Deep learning-based super-resolution in coherent imaging systems. *Sci Rep* **9**, 3926 (2019).

32. Wang, H. *et al.* Early detection and classification of live bacteria using time-lapse coherent imaging and deep learning. *Light Sci Appl* **9**, 118 (2020).

33. Kim, G. *et al.* Rapid label-free identification of pathogenic bacteria species from a minute quantity exploiting three-dimensional quantitative phase imaging and artificial neural network. 596486 Preprint at https://doi.org/10.1101/596486 (2021).

34. Işıl, Ç. *et al.* Phenotypic Analysis of Microalgae Populations Using Label-Free Imaging Flow Cytometry and Deep Learning. *ACS Photonics* **8**, 1232–1242 (2021).

35. Butola, A. *et al.* High spatially sensitive quantitative phase imaging assisted with deep neural network for classification of human spermatozoa under stressed condition. *Sci Rep* **10**, 13118 (2020).

36. Kim, G., Jo, Y., Cho, H., Min, H. & Park, Y. Learning-based screening of hematologic disorders using quantitative phase imaging of individual red blood cells. *Biosensors and Bioelectronics* **123**, 69–76 (2019).

37. Shu, X. *et al.* Artificial-Intelligence-Enabled Reagent-Free Imaging Hematology Analyzer. *Advanced Intelligent Systems* **3**, 2000277 (2021).

38. Javidi, B. *et al.* Sickle cell disease diagnosis based on spatio-temporal cell dynamics analysis using 3D printed shearing digital holographic microscopy. *Opt. Express, OE* **26**, 13614–13627 (2018).

39. O'Connor, T., Anand, A., Andemariam, B. & Javidi, B. Deep learning-based cell identification and disease diagnosis using spatio-temporal cellular dynamics in compact digital holographic microscopy. *Biomed. Opt. Express, BOE* **11**, 4491–4508 (2020).

40. Goswami, N. *et al.* Label-free SARS-CoV-2 detection and classification using phase imaging with computational specificity. *Light Sci Appl* **10**, 176 (2021).

41. O'Connor, T., Shen, J.-B., Liang, B. T. & Javidi, B. Digital holographic deep learning of red blood cells for field-portable, rapid COVID-19 screening. *Opt. Lett., OL* **46**, 2344–2347 (2021).

42. Li, Y. *et al. Deep Learning-enabled Detection and Classification of Bacterial Colonies using a Thin Film Transistor (TFT) Image Sensor.* http://arxiv.org/abs/2205.03549 (2022) doi:10.48550/arXiv.2205.03549.

43. Lin, L.-C., Kuo, Y.-C. & Chou, C.-J. Anti-Herpes Simplex Virus Type-1 Flavonoids and a New Flavanone from the Root of Limonium sinense. *Planta Med* **66**, 333–336 (2000).

44. Kirsi, J. J. *et al.* Broad-spectrum antiviral activity of 2-beta-D-ribofuranosylselenazole-4-carboxamide, a new antiviral agent. *Antimicrobial Agents and Chemotherapy* **24**, 353–361 (1983).

45. Baker, D. A. & Glasgow, L. A. Rapid Plaque Assay for Encephalomyocarditis Virus. *Appl Microbiol* **18**, 932–934 (1969).

46. Zhang, Y. *et al.* Motility-based label-free detection of parasites in bodily fluids using holographic speckle analysis and deep learning. *Light Sci Appl* **7**, 108 (2018).

47. Luo, W., Zhang, Y., Feizi, A., Göröcs, Z. & Ozcan, A. Pixel super-resolution using wavelength

scanning. *Light Sci Appl* **5**, e16060–e16060 (2016).

48. Abdelmageed, A. A. & Ferran, M. C. The Propagation, Quantification, and Storage of Vesicular Stomatitis Virus. *Curr Protoc Microbiol* **58**, e110 (2020).

49. Mudanyali, O. *et al.* Compact, light-weight and cost-effective microscope based on lensless incoherent holography for telemedicine applications. *Lab Chip* **10**, 1417–1428 (2010).

50. Bishara, W., Zhu, H. & Ozcan, A. Holographic opto-fluidic microscopy. *Opt. Express, OE* **18**, 27499–27510 (2010).

51. Greenbaum, A. *et al.* Imaging without lenses: achievements and remaining challenges of wide-field on-chip microscopy. *Nat Methods* **9**, 889–895 (2012).

52. Tseng, D. *et al.* Lensfree microscopy on a cellphone. *Lab Chip* **10**, 1787–1792 (2010).

53. Greenbaum, A. *et al.* Wide-field computational imaging of pathology slides using lens-free on-chip microscopy. *Science Translational Medicine* **6**, 267ra175-267ra175 (2014).

54. Göröcs, Z. & Ozcan, A. On-Chip Biomedical Imaging. *IEEE Rev Biomed Eng* **6**, 29–46 (2013).

55. Wu, Y. & Ozcan, A. Lensless digital holographic microscopy and its applications in biomedicine and environmental monitoring. *Methods* **136**, 4–16 (2018).

56. Zhang, Y., Wang, H., Wu, Y., Tamamitsu, M. & Ozcan, A. Edge sparsity criterion for robust holographic autofocusing. *Opt. Lett., OL* **42**, 3824–3827 (2017).

57. Estimate displacement field that aligns two 2-D or 3-D images - MATLAB imregdemons. https://www.mathworks.com/help/images/ref/imregdemons.html.

58. Reitberger, G. & Sauer, T. Background Subtraction using Adaptive Singular Value Decomposition. *J Math Imaging Vis* **62**, 1159–1172 (2020).

59. Huang, G., Liu, Z., Pleiss, G., Van Der Maaten, L. & Weinberger, K. Convolutional Networks with Dense Connectivity. *IEEE Transactions on Pattern Analysis and Machine Intelligence* 1–1

(2019) doi:10.1109/TPAMI.2019.2918284.

60. Qiu, Z., Yao, T. & Mei, T. Learning Spatio-Temporal Representation With Pseudo-3D Residual Networks. in *Proceedings of the IEEE International Conference on Computer Vision* 5533–5541 (2017).24